# ESCAPE

**Preparing Forecasting Systems for the Next generation of Supercomputers**

# D2.3
# Public release of Atlas under an open source license, which is accelerator enabled and has improved interoperability features

Dissemination Level: public

This project has received funding from the European Union's Horizon 2020 research and innovation programme under grant agreement No 67162

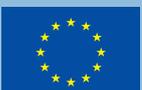

Funded by the
European Union

Co-ordinated by 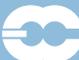 ECMWF

# ESCAPE

**Energy-efficient Scalable Algorithms**
**for Weather Prediction at Exascale**

Author **Willem Deconinck**

Date **31/12/2017**

# Contents

# 1 Executive Summary

In previous deliverable report D1.3 [1], we presented Atlas, a new software library that is currently being developed at the European Centre for Medium-Range Weather Forecasts (ECMWF), with the scope of handling data structures required for numerical weather prediction (NWP) and climate applications in a flexible and massively parallel way. Within the ESCAPE project, Atlas provides a common foundation for many of the Weather & Climate Dwarfs, by supporting their data structure and parallelisation requirements. These requirements are quite versatile between dwarfs or even between prototype implementations of the same dwarf. Some dwarfs are implemented on limited area grids whereas other dwarfs work on global Earth grids. Some dwarfs require elaborate unstructured mesh data structures whereas other dwarfs require a structured grid to underly the numerical algorithms. Of particular relevance for emerging HPC architectures is the support of multiple memory spaces. Fields and data structures require then to be synchronised between memory of a host CPU, and memory of an accelerator (e.g. GPU). Atlas also provides in this regard capabilities to support development of efficient NWP and climate applications, thus constituting a step towards affordable exascale high-performance simulations.

With this deliverable, a first version of the Atlas software libraries is publicly released with a permissive open-source license. The software is freely available for download, and contains a user guide and installation instructions.

The Atlas libraries have been carefully designed with the user's perspective in mind. Even though Atlas is mainly coded in C++, an equivalent Fortran interface is presented without additional runtime overhead. The Fortran interfaces are provided to accommodate existing NWP and climate models that typically consist of Fortran subroutines. The mixed Fortran/C++ design enhances interoperability between NWP and climate models and novel data management techniques. Atlas provides interoperability with accelerator hardware, and can serve as foundation to support higher level abstractions as used in domain specific languages.

# 2 Introduction

## 2.1 Background

ESCAPE stands for Energy-efficient Scalable Algorithms for Weather Prediction at Exascale. The project will develop world-class, extreme-scale computing capabilities for European operational numerical weather prediction and future climate models. ESCAPE addresses the ETP4HPC Strategic Research Agenda 'Energy and resiliency' priority topic, developing a holistic understanding of energy-efficiency for extreme-scale applications using heterogeneous architectures, accelerators and special compute units by:

- Defining and encapsulating the fundamental algorithmic building blocks underlying weather and climate computing;

- Combining cutting-edge research on algorithm development for use in extreme-scale, high-performance computing applications, minimising time- and cost-to-solution;

- Synthesising the complementary skills of leading weather forecasting consortia, university research, high-performance computing centres, and innovative hardware companies.

ESCAPE is funded by the European Commission's Horizon 2020 funding framework under the Future and Emerging Technologies - High-Performance Computing call for research and innovation actions issued in 2014.

## 2.2 Scope of this deliverable

### 2.2.1 Objectives of this deliverable

The *Atlas* library is a software library being developed at ECMWF in the context of its Scalability Programme. As such, at the initiation of ESCAPE, the library was already in a functional state to support the development of the existing dwarfs (see Deliverable D1.1 [2]). It was however still in an early development stage.

During the recently delivered Deliverable D1.3 [1], an internal Atlas version has been released to ESCAPE partners. This deliverable aims at providing an established first open-source release of the *Atlas* library, available to the public. The released version offers interoperability between languages (C/C++ and Fortran), and between CPU and accelerator hardware (NVIDIA GPU's or Intel Xeon Phi).

Further ESCAPE developments also include the application of a Domain Specific Language (DSL) to several dwarfs. The DSL can have different backends, each capable of executing algorithms on different HPC hardware architectures (CPU, GPU, MIC). Especially GPU architectures are very different in nature and algorithms may require copying data back and forth from a host architecture (CPU) to a device (GPU) where computations on the data are performed (see Deliverable D2.4 [3]). The delivered *Atlas* release therefore also includes a new advanced data-storage facility that accommodates host-device synchronisation capabilities with different backends. In practice the GPU backend is currently implemented only for GPU's programmable with the CUDA language (NVIDIA) [4].

### 2.2.2 Work performed on this deliverable

As suggested in Section 2.2.1, a first *Atlas* version was recently released for internal use by ESCAPE partners (Deliverable D1.3 [1]). The majority of the work performed from ESCAPE's initiation to the Deliverable D1.3 has been to design and implement new capabilities as well as redesign and reimplement existing capabilities to accomodate new or evolving requirements.

During the months between Deliverable D1.3 [1] and current deliverable, work has been performed in 3 key areas:

1. Improving user experience.

2. Porting to various compilers and hardware architectures.

3. Improving interoperability with GPU in support of porting Dwarfs to GPU.

4. Implementing new Atlas features, e.g. to interoperate between structured/unstructured grids at multiple resolutions.

5. Devising a strategy to publicly release the Atlas libraries.

As part of improving the user experience (item 1), we paid special attention to the Fortran interfaces. In order to have an interface as close as possible to the C++ Object Oriented design, relatively "new" Fortran 2003 features were used. It has proven that even 15 years after this released Fortran standard, almost all tested compilers had at least a few issues. We reported 9 compiler bugs to PGI, 3 compiler bugs to Intel, 3 compiler bugs to Cray, and encountered 2 compiler bugs for GNU which had already been reported by others previously.

Even though we encountered these compiler bugs, we managed to find a workaround in most cases by first introspecting the compiler capabilities during a configuration phase, and then take different compilation decisions depending on the outcome. This way we managed to increase portability between various compilers and hardware architectures (item 2).

With respect to interoperability with GPU architectures (item 3), we have taken prototype implementations from Deliverable D1.3 [1] further and matured user interfaces. More specifically, interoperability between CUDA enabled memory managed with C++ implementations and Fortran algorithms using OpenACC directives has greatly improved. Additional work was done to enable parallel communications between GPU devices directly without a synchronisation step through the CPU host. The work described by this paragraph has been performed in close collaboration between ECMWF and MeteoSwiss.

Further improvements to Atlas have been implemented during the past months related to interoperability between structured and unstructured grids at different resolutions (item 4). This work is especially useful when integrating multiple Dwarfs together that may rely on different numerical methods using different data structures. Extra care has been taken to minimise the costs of parallel communication.

As for the task of releasing the Atlas libraries with an open-source license, and making it publicly available, there were various decisions to be made: which license, where to host the software, and how to manage contributions from external public collaborators. We have chosen to release the Atlas libraries using the permissive open source Apache-2 license (`http://www.apache.org/licenses/LICENSE-2.0`) to maximise potential use in the community's likely closed-source software. To make the software easily accessible we will host a mirror of our released versions on Github (`http://www.github.com/ecmwf`), which offers as well a way to monitor developments of external developers, that can then easily be added as contributions to subsequent releases. We also provide a user documentation that can be downloaded on the *Atlas* website: `https://software.ecmwf.int/wiki/display/atlas`.

The same argument of creating tools for the community's benefit, used to endorse the development of Atlas as a separate library, has been used to move more common utility features like logging, configuration, smart pointers, etc. from Atlas to separate C++ and Fortran support libraries. These features are deemed so useful that they could benefit a wider software development community than that of numerical weather prediction and climate modelling. Naturally, these required support libraries will be released and managed in the same way.

Atlas also optionally relies for its accelerator support on the open-source GridTools library developed by ESCAPE partner MeteoSwiss and the Swiss national super-computing centre (CSCS). We have greatly improved the installation procedure of Atlas using GridTools, by contributing changes to the build system of both GridTools and Atlas in order to obtain a more uniform and integrated environment. GridTools is not managed by ECMWF, and not is not yet publicly released. It will however be publicly available on Github (`http://www.github.com/GridTools`) by the end of the ESCAPE project, also with a permissive open-source license.

### 2.2.3 Deviations and counter measures

Even though *Atlas* now fully supports mesh generation for regional grids as required by the majority of limited-area models, there is more work that can be done to support other aspects in *Atlas* such as mathematical operators (gradient, divergence, curl) taking into account the used projections of a regional grid. Although this work is ongoing during the course of the ESCAPE project, it is not foreseen as a critical requirement at this moment to develop algorithms relying on *Atlas* for limited area modelling purposes.

It is our aim to support asynchronous parallel communications, which aids in avoiding latency costs by overlapping communications and computations. We initially planned to add support for this feature within the current deliverable, but there were other priorities that took precedence and required more time than expected. One such priority was ensuring portability between multiple compilers and architectures by finding workarounds for Fortran compiler bugs. We still aim to support asynchronous parallel communication, and will continue to work on this. The feature will be added with a subsequent release in the near future. To keep track of the remaining work, it has been added as JIRA tasks in the ESCAPE software collaboration platform [5]. ESCAPE partners will be kept up to date as new features become available in the meantime. This strategy has shown to work effectively over the course of ESCAPE so far.

## 2.3 *Atlas*, a library for numerical weather prediction and climate modelling

The algorithms underlying numerical weather prediction (NWP) and climate models that have been developed in the past few decades face an increasing challenge caused by the paradigm shift imposed by hardware vendors towards more energy-efficient devices. This is because the Dennard scaling (constant power consumption with increasing transistor density) has ended for traditional CPU cores. Rather than increasing clock speeds of the chips, performance is increased by adding more chips. In order to provide a sustainable path to exascale High

Performance Computing (HPC), applications become increasingly restricted by energy consumption. As a result, the emerging diverse and complex hardware solutions have a large impact on the programming models traditionally used in NWP software, triggering a rethink of design choices for future massively parallel software frameworks. In this deliverable report, we present *Atlas*, a new software library that is currently being developed at the European Centre for Medium-Range Weather Forecasts (ECMWF), with the scope of handling data structures required for NWP applications in a flexible and massively parallel way. *Atlas* provides a versatile framework for the future development of efficient NWP and climate applications on emerging HPC architectures. The applications range from full Earth system models, to specific tools required for post-processing weather forecast products. The *Atlas* library thus constitutes a step towards affordable exascale high-performance simulations by providing the necessary abstractions that facilitate the application in heterogeneous HPC environments by promoting the co-design of NWP algorithms with the underlying hardware.

*Atlas* provides data structures for building various numerical strategies to solve equations on the sphere or limited area's on the sphere. These data structures may contain a distribution of points (grid) and, possibly, a composition of elements (mesh), required to implement the numerical operations required. *Atlas* can also represent a given field within a specific spatial projection. *Atlas* is capable of mapping fields between different grids as part of pre- and post-processing stages or as part of coupling processes whose respective fields are discretised on different grids or meshes. The latter is particularly relevant for the physical parametrisations, where some physical processes such as radiation may be represented on a coarser grid or mesh and may need to be projected onto a finer grid or mesh.

The key concepts in the design of the *Atlas* data structure are:

- *Grid*: ordered list of points (coordinates) without connectivity rules;

- *Mesh*: collection of elements linking the grid points by specific connectivity rules;

- *Field*: array of discrete values representing a given quantity;

- *FunctionSpace*: discretisation space in which a *field* is defined.

These concepts are depicted in Figure 1, where we used the sphere to represent a global grid, mesh and field. A *grid* is merely a predefined list of two-dimensional points, typically structured and using two indices `i` and `j` so that point coordinates and computational stencils (for e.g. derivatives) are easily retrieved without connectivity rules. For models using a structured grid point approach, a *grid* is enough to define *fields* with appropriate indexing mechanisms. For element-based numerical methods (generally unstructured) however, the *mesh* concept is introduced that describes connectivity lists linking elements, edges and nodes.

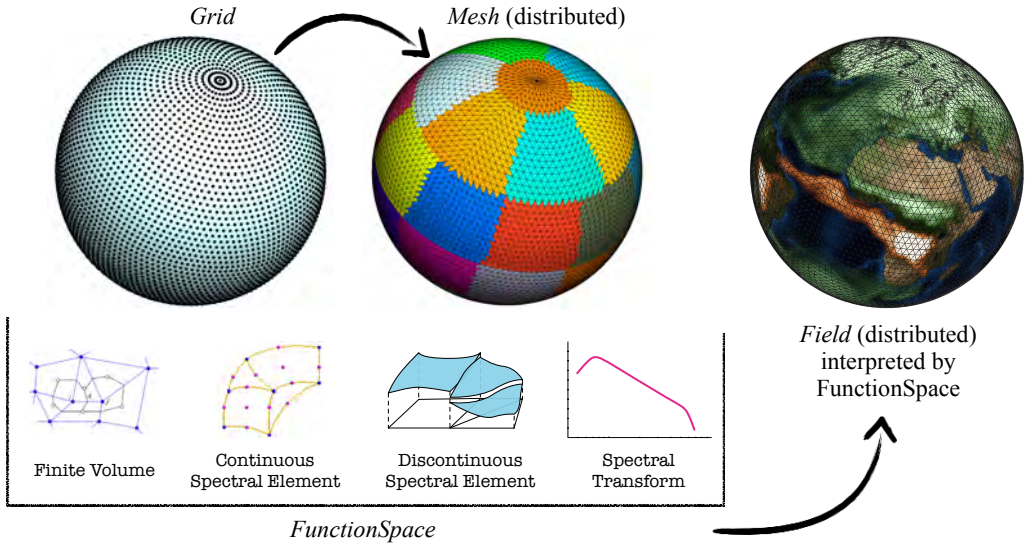

Figure 1: The conceptual design of *Atlas*.

A *mesh* may be decomposed in partitions and distributed among MPI tasks. Every MPI task then allows computations on one such partition. Overlap regions (or halo's) between partitions can be constructed to enable stencil operations in a parallel context.

In addition to these two features, it is necessary to introduce the concept of a *field*, intended as a container of values of a given variable. A *field* can be discretised in various ways. The concept responsible to interpret/provide the discretisation of a *field* in terms of spatial projection (e.g. grid-points, mesh-nodes, mesh-cell-centres) or spectral coefficients is the *function space*. The *function space* also implements parallel communication operations responsible for performing synchronisation of fields across overlap regions, which we refer to as halo-exchange hereafter.

A possible *Atlas* workflow consisting of the creation and discretisation of a *field*, is illustrated in Figure 2, where we also emphasise some additional characteristics of each step. The building blocks illustrated in Figure 2 can then be used to implement additional operations required for specific applications. *Atlas* supplies certain mathematical operations as ready solutions to be plugged in to user software. These operations vary from the computation of gradient, divergence, curl and Laplacian operations to remapping or interpolation of fields defined on different grids.

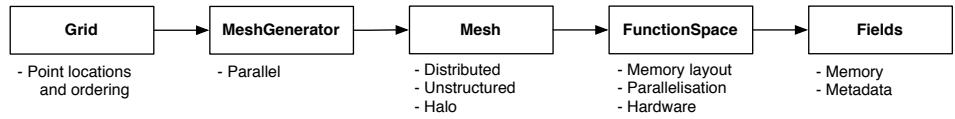

Figure 2: Workflow of *Atlas* starting from *Grid* to the creation of a *Field*, discretised on a *Mesh* and managed by a *FunctionSpace*.

For more comprehensive description on *Atlas* please refer to (Deconinck et. al., 2017) [6], which is published for ESCAPE with open access.

# 3  *Atlas* public release

This section is intended to be a general introduction on how to download, install and run *Atlas*. In particular, in section 3.1 we will present the general system requirements before building the library. Section section 3.2 details how to download the *Atlas* libraries. In section 3.3 we will first describe how to install the internal dependencies required by *Atlas* (if supported by ECMWF) and successively we will outline how to install *Atlas*. Section 3.4 then explains how to check the installation. Finally, in section 3.5 we show how to incorporate *Atlas* in your own software by creating a simple example that initialises and finalises the library.

## 3.1  System requirements

The system requirements for *Atlas* can be summarised as follows:

- **POSIX**: The operating system must be POSIX compliant. Currently this limits the use to UNIX, Linux, and MacOSX operating systems.

- **C++11, Fortran 2008** : *Atlas* uses the programming languages C++ and optionally Fortran. The required standards for these languages are respectively C++11 and Fortran 2008.

- **OpenMP** for C++ (optional): In order for *Atlas* to optionally be able to take advantage of OpenMP multi-threading, the C++ compiler is required to support OpenMP version 3.

- **MPI** for C (optional): To use *Atlas* in a distributed memory application, the system needs to have the MPI libraries for the C-language available.

- **Git**: Required for project management and to download *Atlas*. For use and installation see https://git-scm.com/

- **CMake**: The compilation or build system of *Atlas* is based on CMake 3.3 or higher, which is required to be present on the system. For use and installation see http://www.cmake.org/ .

- **Python**: Required for certain components of the build system. For use and installation see https://www.python.org/. (Known to work with version 2.7.12)

- **CGAL** (optional): *Atlas* can optionally generate meshes on the sphere from an unstructured point cloud. To enable this feature, the Computational Geometry Algorithms Library (CGAL) is required to be present on the system. For use and installation see http://www.cgal.org/.

- **GridTools storage module** (optional): *Atlas* can also optionally make use of the GridTools storage module to support use on accelerator hardware. A requirement here is also the Boost C++ library. When intended for a GPU accelerator, an additional requirement is also that CUDA 6.0 or greater be installed on the system. For installation and download instructions, see <http://www.github.com/gridtools>

## 3.2 Downloading to Atlas

The Atlas libraries can be downloaded via our Github portal <http://www.github.com/ecmwf>. As mentioned in the introduction, some Atlas functionality has been separated into externally versioned software repositories with the intent to modularise common functionality not specific to NWP or climate applications. Following software repositories can be found on the Github portal:

- **ecbuild**: It implements some CMake macros that are useful for configuring and compiling *Atlas* and the other libraries required by *Atlas*. For further information, please visit: <https://software.ecmwf.int/wiki/display/ecBuild>.

- **eckit**: It implements useful C++ functionalities widely used in ECMWF C++ projects. For further information, please visit: <https://software.ecmwf.int/wiki/display/ecKit>

- **fckit** (optional): It implements useful Fortran modules that improve object-oriented programming capabilities or interface the eckit library. For further information, please visit: <https://software.ecmwf.int/wiki/display/fckit>

- **atlas**: Contains the *Atlas* C++ and Fortran libraries. For further information, including comprehensive user documentation please visit: <https://software.ecmwf.int/wiki/display/atlas>

To download *Atlas* and its internal dependencies, following instructions may be used on the command line assuming you have a registered and configured github account:

```
export SRC=$(pwd)/source
mkdir -p ${SRC}
cd ${SRC}

git clone -b master git@github.com:ecmwf/ecbuild
git clone -b master git@github.com:ecmwf/eckit
```

```
git clone -b master git@github.com:ecmwf/fckit
git clone -b master git@github.com:ecmwf/atlas
```

If you don't have a github account, you may still download the repositories instead as archives and extract them in the `${SRC}` directory. Download links are respectively:

- https://github.com/ecmwf/ecbuild/archive/master.zip
- https://github.com/ecmwf/eckit/archive/master.zip
- https://github.com/ecmwf/fckit/archive/master.zip
- https://github.com/ecmwf/atlas/archive/master.zip

The listed software repositories (ecbuild, eckit, fckit, atlas) are all publicly released with this deliverable under the permissive open-source Apache-2.0 license (http://www.apache.org/licenses/LICENSE-2.0).

## 3.3    Compilation and Installation of Atlas

In the following we will outline how to build and install *Atlas* and each of the projects *Atlas* depends on that are not covered by the system requirements. The first step is to create a folder where to build and install each project, and to choose a compilation optimisation level. The following three optimisation levels are recommended:

- `DEBUG`: No optimisation - used for debugging or development purposes only. This option may enable additional bounds checking.

- `BIT`: Maximum optimisation while remaining bit-reproducible.

- `RELEASE`: Maximum optimisation. For some algorithms and using some compilers, too aggressive optimisation can lead to wrong results.

```
export BUILD=$(pwd)/build
export INSTALL=$(pwd)/install
export BUILD_TYPE=BIT
export PATH=${PATH}:${SRC}/ecbuild/bin
```

```
mkdir -p ${BUILD}/eckit;  cd ${BUILD}/eckit
ecbuild --build=${BUILD_TYPE} --prefix=${INSTALL}/eckit -- ${SRC}/eckit
make -j8 install
```

```
mkdir -p ${BUILD}/fckit;  cd ${BUILD}/fckit
ecbuild --build=${BUILD_TYPE} --prefix=${INSTALL}/fckit -- \
  -DECKIT_PATH=${INSTALL}/eckit \
  ${SRC}/fckit
make -j8 install
```

```
mkdir -p ${BUILD}/atlas;  cd ${BUILD}/atlas
ecbuild --build=${BUILD_TYPE} --prefix=${INSTALL}/atlas -- \
  -DECKIT_PATH=${INSTALL}/eckit \
  -DFCKIT_PATH=${INSTALL}/fckit \
  ${SRC}/atlas
```

The following extra flags may be added to Atlas configuration step to fine-tune features

- `-DENABLE_OMP=OFF` — Enable/Disable OpenMP

- `-DENABLE_FORTRAN=OFF` — Disable Compilation of Fortran bindings

- `-DENABLE_BOUNDSCHECKING=ON` — Enable boundschecking in C++ code when indexing arrays. By default BOUNDSCHECKING is ON when the build-type is DEBUG, otherwise the default is OFF.

**Note**

By default compilation is done using shared libraries. Some systems have linking problems with static libraries that have not been compiled with the flag `-fPIC`. In this case, also compile Atlas using static linking, by adding to the ecbuild step for each project the flag: `--static`

**Note**

The build system for the entire software stack presented above is based on ecbuild which facilitates portability across multiple platforms. However some platforms (like ECMWF's HPC) may have a non-standard configuration (in terms of CMake). For these cases ecbuild has a *toolchain* option, which allows you to provide a custom set of rules for a specific platform. The reader is referred to the ecbuild documentation, and the ecbuild "help" : `ecbuild --help`

The building and installation of *Atlas* should now be complete and you can start using it. With this purpose, in the next section we show a simple example on how to create a simple program to initialise and finalise the library.

## 3.4 Inspecting your *Atlas* installation

Once installation of Atlas is complete, an executable called "atlas" can be found in `${INSTALL}/bin/atlas`. Example use is listed:

```
>>> ${INSTALL}/bin/atlas --version
0.10.0

>>> ${INSTALL}/bin/atlas --git
8963cb48

>>> ${INSTALL}/bin/atlas --info
atlas version (0.10.0), git (8963cb48)

  Build:
    build type    : Release
    timestamp     : 20160215122606
    op. system    : Darwin-14.5.0 (macosx.64)
    processor     : x86_64
    c compiler    : Clang 7.0.2.7000181
      flags       :  -O3 -DNDEBUG
    c++ compiler  : Clang 7.0.2.7000181
      flags       :  -O3 -DNDEBUG
    fortran compiler: GNU 5.2.0
      flags       :  -fno-openmp -O3 -funroll-all-loops -finline-functions

  Features:
    Fortran       : ON
    MPI           : ON
    OpenMP        : OFF
    BoundsChecking : OFF
    ArrayDataStore : GridTools
    GPU           : OFF
    Trans         : OFF
    Tesselation   : ON
    gidx_t        : 64 bit integer

  Dependencies:
    eckit version  (0.12.3), git (7b76818e)
    fckit version  (0.3.1), git (9ba8d82)
```

This executable gives you information respectively on the version, a more detailed git-version-controlled identifier, and finally a more complete view on all the features that Atlas has been compiled with, as well as compiler and compile flag information. Also printed are the versions of used dependencies such as eckit and fckit.

## 3.5 Using *Atlas* in your project

In this section, we provide a simple example on how to link *Atlas* in your own software. We will show a simple "Hello world" program that initialises and finalises the library, and uses the internal *Atlas* logging facilities to print "Hello world!".

Note that *Atlas* supports both C++ and Fortran. Therefore, we will show equivalent examples using both C++ and Fortran.

```cpp
// file: hello-world.cc

#include "atlas/library/Library.h"
#include "atlas/runtime/Log.h"

int main(int argc, char** argv)
{
    atlas::Library::instance().initialise(argc, argv);
    atlas::Log::info() << "Hello world!" << std::endl;
    atlas::Library::instance().finalise();

    return 0;
}
```

Listing 1: Using *Atlas* in a C++ project

```fortran
! file: hello-world.F90

program hello_world

use atlas_module, only : atlas_library, atlas_log

call atlas_library%initialise()
call atlas_log%info("Hello world!")
call atlas_library%finalise()

end program
```

Listing 2: Using *Atlas* in a Fortran project

First, the *Atlas* library is initialised. In C++ this function requires two arguments `argc` and `argv` from the command-line. In Fortran these arguments are automatically provided by the Fortran runtime environment. This function is used to set up the logging facility and for the initialisation of MPI (Message Passage Interface).

Following initialisation, we log "Hello world!" to the `info` channel. *Atlas* provides

4 different log channels which can be configured separately: `debug`, `info`, `warning`, and `error`. By default all log channels print to the std::cout stream, and the debug channel can be switched on or off (irrespective of compilation with the DEBUG build type) by setting the environment variable `ATLAS_DEBUG=1` or `ATLAS_DEBUG=0`. Not specifying `ATLAS_DEBUG` is treated as `ATLAS_DEBUG=0`. Finally we end the program after finalising the *Atlas* library.

>
> **Note**
>
> The logging facility exposed by *Atlas* is implemented by *eckit*. The Fortran interface is using *fckit*, which also delegates its implementation to *eckit*. For this reason, logging through C++ or Fortran shares the same infrastructure, which ensures that the logging is consistent in mixed C++/Fortran codes.

### Standard code compilation

Compiling the C++ example with the GNU C++ compiler:

```
g++ hello-world.cc -o hello-world \
   $(pkg-config ${INSTALL}/atlas/lib/pkgconfig/atlas.pc --libs --cflags)
```

Compiling the Fortran example with the GNU Fortran compiler:

```
gfortran hello-world.F90 -o hello-world \
   $(pkg-config ${INSTALL}/atlas/lib/pkgconfig/atlas.pc --libs --cflags)
```

We can now run the executable:

```
>>> ./hello-world
Hello world!
```

We can run the same executable with debug output printed during Atlas initialisation:

```
>>> ATLAS_DEBUG=1  ./hello-world
```

The output now shows in addition to `Hello world!` also some information such as the version of *Atlas* we are running, the identifier of the commit and the path of the executable, similarly to the output of `atlas --info` in Section 3.4.

## Code compilation using ecbuild

As *Atlas* is an ecbuild (CMake) project, it integrates easily in other ecbuild (CMake) projects. Two sample ecbuild projects are shown here that compile the "hello-world" example code, for respectively the C++ and the Fortran version.

An example C++ ecbuild project would look like this:

```
1  # File: CMakeLists.txt
2  cmake_minimum_required(VERSION 3.3.2 FATAL_ERROR)
3  project(hello_world CXX)
4
5  include(ecbuild_system NO_POLICY_SCOPE)
6  ecbuild_requires_macro_version(2.7)
7  ecbuild_declare_project()
8  ecbuild_use_package(PROJECT atlas REQUIRED)
9  ecbuild_add_executable(TARGET   hello-world
10                         SOURCES  hello-world.cc
11                         INCLUDES ${ATLAS_INCLUDE_DIRS}
12                         LIBS     atlas)
13 ecbuild_print_summary()
```

An example Fortran ecbuild project would look like this:

```
1  # File: CMakeLists.txt
2  cmake_minimum_required(VERSION 3.3.2 FATAL_ERROR)
3  project(hello_world Fortran)
4
5  include(ecbuild_system NO_POLICY_SCOPE)
6  ecbuild_requires_macro_version(2.7)
7  ecbuild_declare_project()
8  ecbuild_enable_fortran(MODULE_DIRECTORY ${CMAKE_BINARY_DIR}/module
9                         REQUIRED)
10 ecbuild_use_package(PROJECT atlas REQUIRED)
11 ecbuild_add_executable(TARGET   hello-world
12                         SOURCES hello-world.F90
13                         INCLUDES ${ATLAS_INCLUDE_DIRS}
14                                  ${CMAKE_CURRENT_BINARY_DIR}
15                         LIBS  atlas_f)
16 ecbuild_print_summary()
```

To compile the ecbuild project, you have to first create an out-of-source build directory and point ecbuild to the directory where the CMakeLists.txt is located.

```
mkdir -p build; cd build
ecbuild -DATLAS_PATH=${INSTALL}/atlas ../
make
```

Note that in the above command we needed to provide the path to the *Atlas* library installation. Alternatively, **ATLAS_PATH** may be defined as an environment variable. This completes the compilation of our first example that uses *Atlas* and generates an executable into the bin folder (automatically generated by CMake) inside our builds directory. For more information on using ecbuild, or CMake, see `https://software.ecmwf.int/wiki/display/ECBUILD/ecBuild`.

This completes your first project that uses the *Atlas* library.

## 3.6 Contributing to Atlas

As Atlas and its libraries are licensed under the open-source Apache-2.0 license, you are free to create a fork, and develop additional features currently missing in the library. To merge your contributions back into a subsequent version, you can use Github features to create a "pull request" which will notify the *Atlas* maintainers to review and accept the contributions.

# 4 Atlas interoperability features

## 4.1 Interoperability between programming languages

*Atlas* is primarily written in the C++ programming language. The C++ programming language facilitates OO design and is high performance computing capable. The latter is due to the support C++ brings for hardware specific instructions. In addition, the high compatibility of C++ with C allows *Atlas* to make use of specific programming models such as CUDA to support GPU's, and facilitates the creation of C-Fortran bindings to create generic Fortran interfaces.

With much of the NWP operational software written in Fortran, significant effort in the *Atlas* design has been devoted to having a Fortran OO Application Programming Interface (API) wrapping the C++ concepts as closely as possible.

The Fortran API mirrors the C++ classes with a Fortran derived type, whose only data member is a raw pointer to an instance of the matching C++ class. The Fortran derived type also contains member functions or subroutines that delegate its implementation to matching member functions of the C++ class instance. Since Fortran does not directly interoperate with C++, C interfaces to the C++ class member functions are created first, and it is these interfaces that the Fortran derived type delegates to. The whole interaction procedure is schematically shown in Figure 3. The overhead created by delegating function calls from the Fortran API to a C++ implementation can be disregarded if performed outside of a computational loop. *Atlas* is primarily used to manage the data structure in a OO manner and the actual field data should be accessed from the data structure before a computational loop starts.

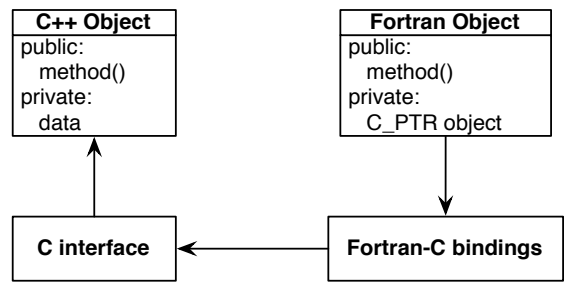

Figure 3: Construction of the Fortran interface to the C++ design. When a method in the Fortran object is called, it will actually be executed by the instance of its matching C++ class, through a C interface.

Because *Atlas* concepts are available both in Fortran and C++, we can create *Atlas* objects and share them between C++ and Fortran routines. This means that developers are not restricted by the language and, depending on the task, they may choose to implement certain capabilities with the language that is best suited.

## 4.2 Interoperability between programming models

When it comes to programming GPU accelerators using Fortran, a natural choice is to use OpenACC directives [7] as portability of code between CPU and GPU can be achieved with minimal changes. *Atlas* GPU-capable data structures are however internally allocated on the GPU using CUDA routines [4]. With the currently released *Atlas* version it is possible to map the CUDA allocated memory in C++ with OpenACC-compatible Fortran arrays. This allows one to have data structures in C++ and use them to develop algorithms using the CUDA programming model, as well as use the same data structures in Fortran and use them to develop algorithms using the OpenACC programming model.

A simple Fortran program is shown in listing 3, illustrating how an *Atlas* field can be modified in Fortran on a GPU.

```fortran
program main
use atlas_module
implicit none

type(atlas_Field)  :: field
real(8), pointer   :: view(:,:)
integer :: i,j
integer, parameter :: Ni = 2
integer, parameter :: Nj = 3

call atlas_library%initialise()
field = atlas_Field(kind=atlas_real(8),shape=[Nj,Ni])
call field%data(view)

!$acc data present(view)
!$acc kernels
do i=1,Ni
  do j=1,Nj
    view(j,i) = i*100 + j
  enddo
enddo
!$acc end kernels
!$acc end data

call CXX_modify_field_using_CUDA(field)
call field%update_host()

call atlas_library%finalise()
end program
```

Listing 3: Fortran interoperability between Atlas and OpenACC

Upon creation of the field, the memory is allocated both on the host CPU and on the GPU device. When requesting access to a view of the data the CUDA allocated memory is mapped to the OpenACC runtime. The "data present(view)" clause then tells OpenACC that the view is already available on the GPU device memory so that no copies are required. After modifying the field on the GPU device using OpenACC we can still modify the field on the GPU by a subroutine implemented in C++ that uses CUDA. Finally we can use *Atlas* functionality to update the field on the host CPU by copying the data back from the GPU device. An example C++ function that could modify the field using CUDA is shown in listing 4.

```cpp
__global__ void cuda_kernel(ArrayView<double, 2> view)
{
  int Nj = view.shape(1);
  for(int j=0; j<Nj; ++j ) {
    view(threadIdx.x,j) += 1000;
  }
}

void modify_field_using_CUDA( atlas::Field field ) {
  auto view =  atlas::array::make_device_view(field);
  cuda_kernel<<<field.shape(0),1>>>( view );
}
```

Listing 4: C++ Interoperability between Atlas and CUDA

## 4.3 Interoperability between multiple grids and resolutions

The Earth system in NWP or climate models is often constituted by a number of model components, where each model component is responsible for a different aspect of the Earth system model. Typical examples are an atmosphere model, ocean model, wave model, radiation model, and a sea-ice model. Each model may be based on different grids with different resolutions and in that case it is a challenge to integrate the different model components together.

Although *Atlas* is not a coupler like OASIS [8] or ESMF [9], it can be seen as a toolkit that provides techniques to remap a field from one grid to another. In order to substantially avoid parallel communication, one grid can be chosen to be a *master* grid (e.g. the fine atmospheric model grid) and its domain-decomposition is well balanced. Subsequent grids (e.g. a coarse radiation model grid) can then be distributed following the domain decomposition of the *master* grid. Figure 4 illustrates how a coarse "F8" grid denoted by solid points is distributed following the domain decomposition of a mesh created for a finer "O32" grid. For a comprehensive description on what these grid names indicate, please refer to ESCAPE Deliverable D1.3 [1] or the Atlas journal article [6].

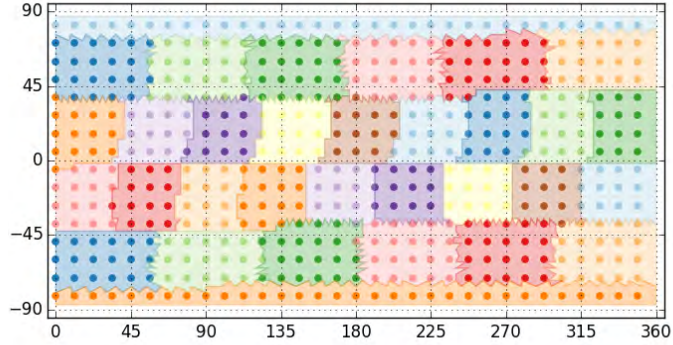

Figure 4: Interoperability between multiple grids or resolutions. Shaded area shows the domain decomposition of a mesh created for a "O32" grid. Solid points denote the gridpoints of a coarser "F8" grid, with colours denoting the domain decomposition. The coarse grid domain decomposition matches the domain decomposition of the finer mesh.

In Figure 5 we lay out a schematic of how to set up two grids of different resolution with matching domain decomposition and then create a remapping operator that can be used to couple distinct earth system components.

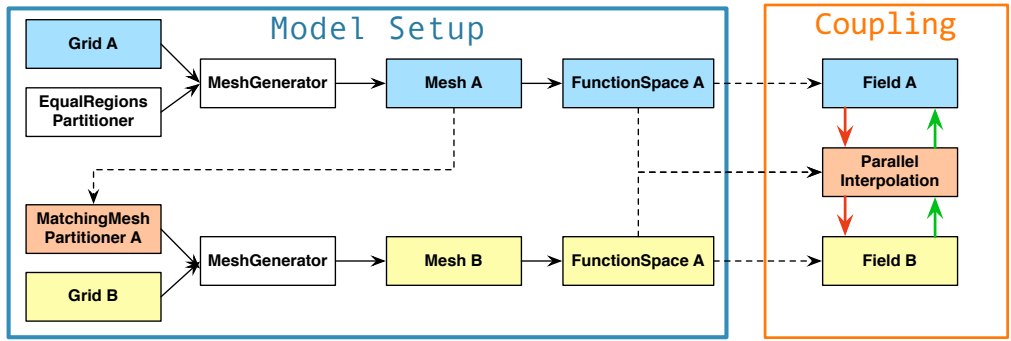

Figure 5: A schematic of how to set up two grids of different resolution with matching domain decomposition, and a remapping operator.

A concrete example of using the described approach is given in Figure 6. An analytically defined field is initialised on a fine "O32" grid and remapped to a coarser "F8" grid. This remapping happened distributed over 32 MPI partitions without any parallel interpolation required. The used interpolation algorithm is a linear element-based algorithm in 3D on the sphere, as described in [10].

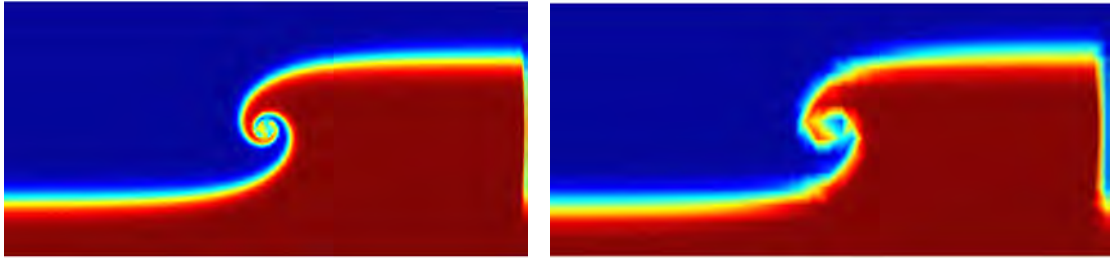

Figure 6: Remapping example of a field initialised on a "O32" grid to a coarser "F8".

# 5 Conclusions

The *Atlas* C++/ Fortran library provides flexible data structures for both structured and unstructured meshes and is intended to be applied in NWP or Climate modelling codes. During the ESCAPE project the *Atlas* library has matured substantially, and gained new capabilities in support of accelerator hardware and responded to requirements for the wider NWP community. In particular the limited area modelling (LAM) community can now use *Atlas* data structures supporting regional grids with projections. Thanks to the ESCAPE project, *Atlas* was scrutinised by experts in software design for accelerator hardware. Also the LAM community was involved at the early development stage of *Atlas*. It would have proved a much more difficult task to redesign *Atlas* after the library would have matured without the involvement of the ESCAPE partners.

With this ESCAPE deliverable, we release *Atlas* publicly with an open-source Apache-2.0 license. The source code of *Atlas* is available on ECMWF's Github space (`http://www.github.com/ecmwf`) with the documentation of *Atlas* on its public website: `https://software.ecmwf.int/wiki/display/atlas`. The public Github software hosting platform will be used to monitor external developments and accept external code contributions.

# Document History

| Version | Author(s) | Date | Changes |
|---------|-----------|------|---------|
| 0.1 | W. Deconinck | 4/12/17 | First draft |
| 0.2 | W. Deconinck | 11/12/17 | Second draft |
| 0.3 | W. Deconinck | 14/12/17 | Version for internal review |
| 0.4 | W. Deconinck | 18/12/17 | Corrections from LU |
| 0.5 | W. Deconinck | 23/12/17 | Corrections from ICHEC |
| 1.0 | W. Deconinck | 28/12/17 | Final version |

# Internal Review History

| Internal Reviewers | Date | Comments |
|--------------------|------|----------|
| LU | 15/12/17 | No comments |
| ICHEC | 20/12/17 | No comments |

# Effort Contributions per Partner

| Partner | Efforts |
|---------|---------|
| ECMWF | 2 pm |
| MeteoSwiss | 5 pm |
| *Total* | 7 pm |

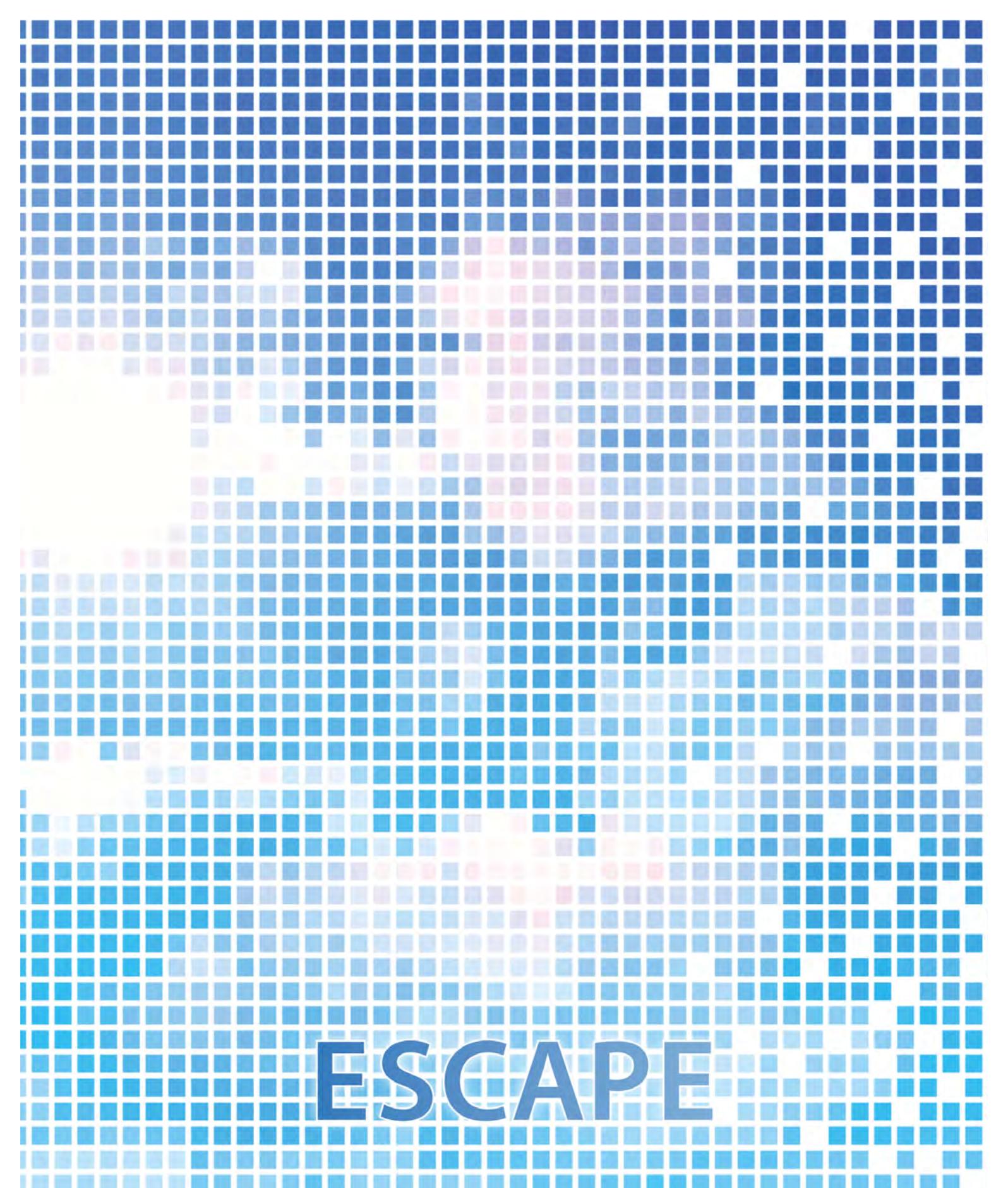